\documentstyle[twocolumn,aps,epsfig]{revtex}
\begin{document}
\title{Frustration driven lattice distortion; an NMR investigation of Y$_{2}$Mo$%
_{2} $O$_{7}$}
\author{Amit Keren$^1$ and Jason S. Gardner$^2$}
\address{$^{1}$Department of Physics, Technion - Israel Institute of Technology,\\
Haifa 32000, Israel.\\
$^{2}$ NRC Canada, NPMR, Chalk River, Laboratories,\\
Chalk River, Ontario, K0J 1J0 Canada.}
\address{{\rm (Received: )}}
\address{\mbox{ }}
\address{\vspace{4mm}\parbox{14cm}{\rm \mbox{ }\mbox{ }\\ We have investigated the $^{89}$Y NMR spectrum and spin lattice relaxation, $%
T_{1}$, in the magnetically frustrated pyrochlore Y$_{2}$Mo$_{2}$O$_{7}$. We
find that upon cooling the spectrum shifts, and broadens asymmetrically. A
detailed examination of the low $T$ spectrum reveals that it is constructed
from multiple peaks, each shifted by a different amount. We argue that this
spectrum is due to discrete lattice distortions, and speculate that these
distortions relieve the frustration and reduce the system's energy. \\}}
\address{\mbox{ }}
\address{\vspace{2mm}\parbox{14cm}{\rm PACS numbers: }}
\maketitle

{\it Geometric frustration}~\cite{toulouse,reviews} occurs in magnetic
systems where competition between magnetic nearest neighbor interactions and
the local symmetry occurs. Geometrically frustrated magnets, such as those
on a kagom\'{e} (2D) or pyrochlore (3D) lattice, often show signs of
spin-glass-like behavior. Examples are the oxide pyrochlore compounds
including Y$_{2}$Mo$_{2}$O$_{7}$~[YMoO] \cite
{dunsiger,gingras_ymoo,gardner_y2mo2o7}, Tb$_{2}$Mo$_{2}$O$_{7}$~\cite
{dunsiger,gaulin}, Y$_{2}$Mn$_{2}$O$_{7}$~\cite{y2mn2o7} and the kagom\'{e}
based systems SrCr$_{8}$Ga$_{4}$O$_{19}$~\cite{MendelsPRL00,SCGO} and Ba$%
_{2} $Sn$_{2}$Ga$_{3}$ZnCr$_{7}$O$_{22}$~\cite{BSGZCO}. Out of these,
perhaps the best studied example is YMoO where detailed analysis of
susceptibility measurements show both scaling and irreversibility \cite
{gingras_ymoo} exactly as expected in conventional, chemically-disordered,
spin glasses \cite{mydosh}. This is very intriguing since it is well
established that glassiness requires BOTH randomness (disorder) and
frustration \cite{BinderPMP86}. In the kagom\'{e} case it was recently shown
that the disordered magnetic sublattice is responsible for the glassiness 
\cite{MendelsPRL00}. However, the pyrochlore compounds are nominally
disorder free. Therefore, the origin of their glassy behavior is still an
open question.

Here we address this question by studying the internal magnetic field
distribution in YMoO using $^{89}$Y NMR. Our major finding is a non-random
distortion of the Mo sub-lattice starting at temperatures as high as $200$%
~K. Such a distortion was previously detected by the EXAFS measurements of
Booth {\it et al.} \cite{BoothPRB00} at $T=15$~K. These distortions produce
many non-equivalent $^{89}$Y sites and result in well separated, regularly
spaced $^{89}$Y NMR peaks. We argue that such distortions can relieve the
frustration and lower the systems energy, consequently resulting in various
values of the spin-spin coupling constant, $J$, and the observed
spin-glass-like behavior.

\begin{figure}[h,t]
\begin{center}
\mbox{\epsfig{file=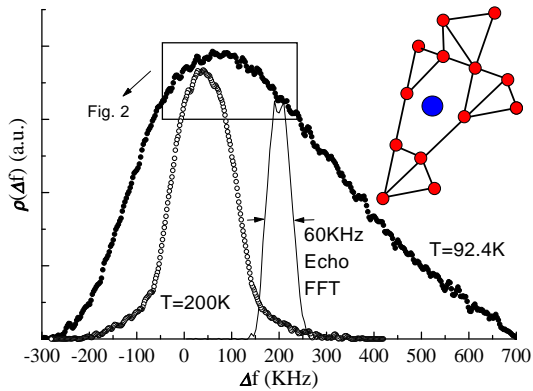,width=90mm}}
\end{center}                         
\caption[]{The spectrum of the $^{89}$Y line at two temperatures. The units
of the abscissa are arbitrary (for each $T$) and are chosen for presentation
purposes only. The marked area is replotted in Fig.~\ref{Zoom}. The inset
shows the local environment of the $^{89}$Y in Y$_{2}$Mo$_{2}$O$_{7}$. }
\label{FullRangeSweep}
\end{figure}

In YMoO, the magnetic Mo-ions and the non-magnetic Y-ions form two
sub-lattices of corner-sharing tetrahedra. The sub-lattices interpenetrate
in a way in which the Y is in the center of a Mo hexagon; pairs of Mo ions
on this hexagon form the edges of the corner sharing tetrahedra. The Y
environment is shown in the inset of Fig.~\ref{FullRangeSweep}.
Susceptibility measurements show that the spin-1, Mo$^{4+}$ ions have an
effective moment of 2.55$\mu _{B}$. These moments interact
antiferromagnetically, giving rise to a Curie-Weiss temperature $\theta_{%
\text{cw}}=200$~K \cite{RajuPRB92}. The transition to the glassy state
occurs at $T_{g}=22.5$~K, as measured by the irreversibility in field-cooled
and zero-field-cooled magnetisation and is characterized by a strong
suppression of spin fluctuations below this temperature~\cite
{dunsiger,gardner_y2mo2o7}. The preparation of YMoO is described elsewhere~ 
\cite{gardner_y2mo2o7}. Rietveld refinement of neutron powder diffraction
data displays no sign of site exchange between Y and Mo or deviation from
the nominal oxygen stoichiometry (an upper limit of 4\% and 1\% was placed
respectively).

We obtain the NMR spectrum by sweeping the external field $H_{\text{ext}}$
in a constant applied RF frequency $f_{\text{app}}=18.13$~MHz with a 50~G
step size. For each field we apply a $\pi /2$-$\pi $ pulse sequence and
record the entire time-dependent echo. We Fourier transform the recorded
echo and shift the frequency scale by $\Delta f=(^{89}\gamma /2\pi
)(H_{0}-H_{\text{ext}})$; $H_{0}$ is the resonance field at $T=300$~K. We
then add all the Fourier transforms to obtain the distribution of frequency
differences $\rho (\Delta f)$. This method, known as the Fourier Step Sum
(FSS) \cite{ClarkRSI95}, allows one to construct the lines structures with
the frequency resolution of the Fourier transform ($5$~kHz in our case)
rather than on the larger FWHM spectrometer bandwidth ($60$~kHz in our case).

The sweeps at low temperatures were done in $\sim$ 80 different fields over
4kG. These sweeps took several days during which the temperature was kept
constant to within $0.2$~K. For a high $T$ sweep we used 40 different field
values. The integrated intensity (normalized by $T$ due to the nuclear
polarization) is similar in all temperatures indicating that we are
detecting all the nuclear spins in the system. We also measured the spin
lattice relaxation time, $T_{1}$, at different positions in the spectrum to
verify that our repetition rate is much longer than the longest $T_{1}$. At
present our experiment is limited to liquid nitrogen temperatures due to the
large spectral width at low $T$ (to be demonstrated below) which results in
low intensities at each applied field, and the difficulty of using high RF
power below this temperature.

In Fig.~\ref{FullRangeSweep} we plot $\rho (\Delta f)$ at $T=200$ and $92.4$
K using different arbitrary units for each line for presentation purposes.
The high $T$ spectra is nearly a perfect Gaussian except that some structure
is seen at its top. In contrast, the low $T$ spectra is not symmetric and
the spectral weight shifts to high frequencies. In addition, the spectrum
does not appear smooth but rather a combination of many peaks. In Fig.~\ref
{FullRangeSweep} we also present, with the use of a solid line, the FFT of a
single echo obtained at $H_{\text{ext}}=8.515$~T but with the frequency
shifted by $(\gamma /2\pi )(H_{0}-H_{\text{ext}})$. This FFT shows a clear
splitting at its top. This splitting is much larger than the noise of the
FFT which is not observable on the scale of this figure. This FFT also
demonstrates that the FWHM is $60$~kHz.

\begin{figure}[h,t]
\begin{center}
\mbox{\epsfig{file=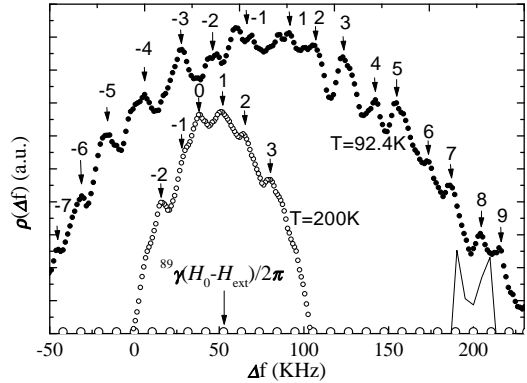,width=90mm}}
\end{center}                         
\caption[]{The same as in Fig.~\ref{FullRangeSweep} but only the center of the
line is shown. Multiple peaks are observed in the spectrum and are indexed.
In the $T=92.4$~K spectrum the index ``0'' is missing.}
\label{Zoom}
\end{figure}

In Fig.~\ref{Zoom} we replot the area marked with a rectangle in Fig.~\ref
{FullRangeSweep} to reveal the rich structure in the spectrum. Clearly, it
is a superposition of many distinct lines, each shifted by a different
amount. The noticeable width of each peak is $\sim 20$ kHz. It is therefore
not surprising that these peaks, although noticeable in a standard field
sweep experiment, are much better resolved using the FSS method. The FFT of
the same single echo shown in Fig.~\ref{FullRangeSweep} is depicted by the
solid line in the right lower corner of Fig.~\ref{Zoom}; the splitting is
much clearer here. Nevertheless, it is important to verify that the presence
of a discrete set of lines is not due to the discrete number of external
fields $H_{\text{ext}}$. Therefore, we present $(^{89}\gamma /2\pi
)(H_{0}-H_{\text{ext}})$, for all the $H_{\text{ext}}$ we have applied, with
open circles on the abscissa. The 16 different NMR lines in the figure are
obtained using 26 different external fields. Clearly, the peaks are
incommensurate with the applied fields and are not a result of the spectrum
construction method.

\begin{figure}[h,t]
\begin{center}
\mbox{\epsfig{file=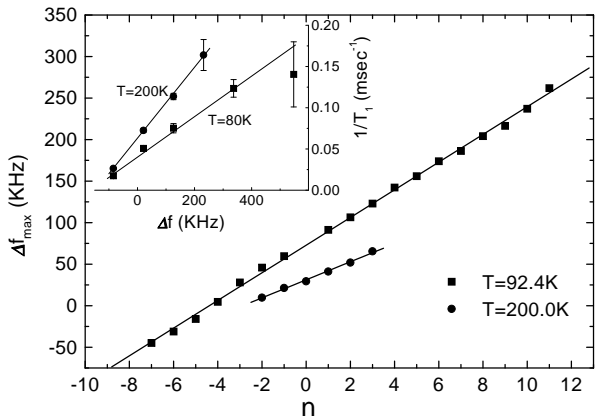,width=90mm}}
\end{center}                         
\caption[]{The frequency differences at which the spectrum peaks $\Delta f_{%
\text{max}}$ is plotted vs. the index of the peak (see Fig.~\ref{Zoom}) for
two different temperatures. A linear fit is shown by the solid lines. The
inset shows the spin lattice relaxation rate, $1/T_{1}$, at different
frequencies in the spectrum and at two different temperatures.}
\label{DeltaFrq}
\end{figure}

In Fig.~\ref{Zoom} we index the frequencies $\Delta f_{\text{max}}$ at which
the peaks are found in an attempt to understand the spectrum. Only a portion
of the indexes are shown. We found that at $T=200$~K, $\Delta f_{\max }(n)$
is a linear function of $n$ as demonstrated in Fig.~\ref{DeltaFrq}. A
similar result is obtained at $T=92.4$~K (here the ``0''th peak is missing).
It is also important to mention that at temperatures higher than 200K, which
is $\theta_{\text{cw}}$, we observed a smooth NMR line. Thus, below $\theta_{%
\text{cw}}$ the NMR line is constructed from a set of resonance frequencies,
and these frequencies are quantized.

Booth {\it et al.} argued that in YMoO the distortion is along the Mo-Mo
bond, and not along the Y-Mo bond \cite{BoothPRB00}. In addition, the
absolute value of the NMR line shift indicates a field of $\sim 1$~kG at the 
$^{89}$Y site, which cannot originate from dipolar coupling. Therefore, it
is reasonable to assume that the dominant coupling between $^{89}$Y and the
Mo-spin is transferred hyperfine and that it is uniform throughout the
sample. This, combined with the observation of multiple peaks, leads us to
the conclusion that in YMoO there is a discrete set of different local
environments, each with its own local spin susceptibility $\chi _{\text{loc}%
} $. The relationship between this susceptibility and the observed frequency
difference $\Delta f$ is given by 
\begin{equation}
\chi _{\text{loc}}\propto \frac{\Delta f}{f}\approx \frac{\Delta f}{f_{\text{%
app}}}.  \label{HypefineChi}
\end{equation}
The presence of a quantized set of NMR frequencies comes about because $\chi
_{\text{loc}}$ has discrete values.

\begin{figure}[h,t]
\begin{center}
\mbox{\epsfig{file=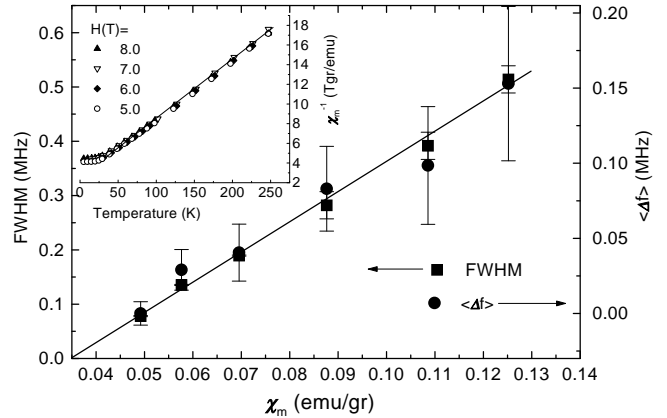,width=90mm}}
\end{center}                         
\caption[]{The mean and FWHM of the frequency difference distribution $\protect%
\rho(\Delta f)$ versus susceptibility. The inset is the inverse macroscopic
DC-susceptibility $H/M$ vs. temperature at several fields between $5$~and $8$%
~T .}
\label{WidthVsT}
\end{figure}

Further information regarding the susceptibility distribution in YMoO was
obtained by comparing the magnetization with the mean and width of the
spectrum. We performed magnetization measurements in fields nearing the ones
used for the NMR measurements. We found that $M$ is a linear function of $H$
at all the relevant temperatures. Therefore, we plot the inverse {\em %
macroscopic } susceptibility $\chi _{m}^{-1}=H/M$ vs temperature in the
inset of Fig.~ \ref{WidthVsT}. The mean $\left\langle \Delta f\right\rangle $
of $\rho (\Delta f)$ is plotted vs. $\chi _{m}$, with the temperature as an
implicit parameter, in Fig.~ \ref{WidthVsT}. The error bars are takes as
10\% of the FWHM. A linear relation is found. In addition, we characterize
the NMR spectrum by the FWHM of $\rho (\Delta f)$, which is also depicted in
Fig.~\ref{WidthVsT} as a function of susceptibility. Again the dependence is
linear, but the extrapolation of the FWHM to zero yields $\chi _{m}=0.035$
emu/gr. If the origin of the line width is frozen disorder (such as Y-Mo
site exchange) we would expect FWHM $\propto \chi _{m}$ (the susceptibility
would be the only temperature dependent parameter). Since this is not the
case, we conclude that the disorder must set in at the temperature where the
FWHM $>0$. From the extrapolation of the susceptibility data this should
occur at $T\sim 430$~K.

Dynamical information regarding the local environment could be obtained by
measurements of the spin lattice relaxation rate $1/T_{1}$ at different
frequencies along the spectrum. The $\Delta f$ dependence of $1/T_{1}$ is
plotted as an inset to Fig.~\ref{DeltaFrq}, at two different temperatures. A
linear dependence of the form $1/T_{1}=n+m\Delta f$ is found with $%
n_{200}=0.062(1)$ and $n_{80}=0.040(1)$ msec$^{-1}$, and $%
m_{200}=4.2(1)\times 10^{-4}$ and $m_{80}=2.6(2)\times 10^{-4}$, where the
subscript is the temperature at which m and n were evaluated. Interestingly,
the average $\left\langle 1/T_{1}\right\rangle =n+m\left\langle \Delta
f\right\rangle $ is $0.08$~msec$^{-1}$ for both temperatures.

\begin{figure}[h,t]
\begin{center}
\mbox{\epsfig{file=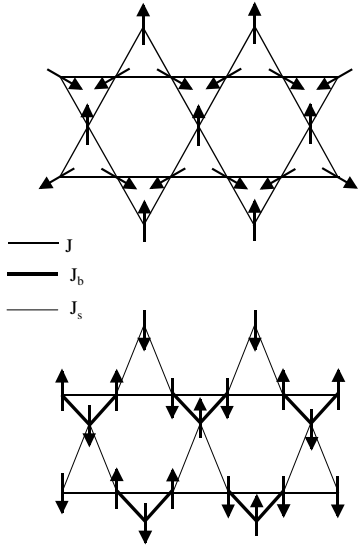,width=60mm}}
\end{center}                         
\caption[]{Schematic demonstrating the concept of frustration induced
distortion for the kagom\'{e} lattice. (a) Before the distortions the
coupling between all spins is $J$. (b) When odd lines move up, the coupling
between spin on diagonal lines is either bigger ($J_{b}$) or smaller ($J_{s}$%
) than $J$ depending on the length of the bond. If $J_{b}>1.5J$ the spin
arrangement in panel (b) has a lower magnetic energy than the one in panel
(a). The system might therefore prefer to distort.}
\label{Distortion}
\end{figure}

Now we would like to discuss the possible origin of the multiple distinct
local environments. The simplest scenario is that they emerge from
impurities, which, in principal could generate alternating magnetic fields
in their vicinity. However, the amount of disorder in the sample, the fact
that it must be temperature dependent, the theoretical expectation of fast
healing (even in the ground state an impurity impacts only its nearest
neighbor) \cite{VillainZPhysB79}, and the experimental observation of short
correlation length ($<0.5$ nm)~\cite{gardner_y2mo2o7} all point away from
this possibility. The other option is lattice distortions. It is conceivable
that such a distortion relieves the frustration and reduces the total system
energy; this was discussed theoretically in detail for classical spins \cite
{TeraoJPSJ95}, quantum spin \cite{Yamashita}, and by numerical simulations 
\cite{Bellier00}. For simplicity we demonstrate the idea of frustration
driven lattice distortion here on the 2-dimensional kagom\'{e} lattice of
corner sharing triangles. In Fig.~\ref{Distortion} we show a kagom\'{e}
lattice where all the bonds are identical. In this case the energy, $E$, per
number of bonds $N$, is $E/N=-J$. We also show a distorted kagom\'{e}
lattice where there are three types of interactions with coupling constants $%
J$, as before, $J_{b}$ (bigger than $J$) for the shorter bonds, and $J_{s}$
(smaller than $J$) for the longer bonds. In this case the main objective of
the spins is to satisfy the shorter bonds. One way of doing this is by
rearranging themselves into a collinear spin configuration as demonstrated
in the figure. In this case, and provided that $J_{s}+J_{b}=2J$, we find $%
E/N=-(2/3)J_{b}$ so that if $J_{b}>(3/2)J$ the lattice has magnetic
``motivation'' to distort.

In the 3D pyrochlore lattice the situation is more complicated since
theoretically a collinear spin configuration belongs to the ground state
manifold. In addition, the reduction of magnetic energy competes with an
increase of the elastic energy due to the distortion, and, at present, it is
not clear that YMoO will benefit energetically from such a distortion. Even
less clear is the reason why the distortion should lead to the discreteness
of sites. However, the recent finding that the two-dimensional $S=1/2$
frustrated antiferromagnet Li$_{2}$VOSiO$_{4}$ system also distorts prior to
its spin freezing \cite{MelziPRL00} is encouraging.

In conclusion we have evidence from $^{89}$Y NMR of multiple discrete values
of the local susceptibility at the Y sites in the spin glass like pyrochlore
Y$_{2}$Mo$_{2}$O$_{7}$. We speculate that this is a result of small discrete
changes in the Mo-Mo bond lengths which are detectable in NMR at $T<200$~K.
We suggest that this distortion relieves the geometric frustration allowing
the system to enter a unique ground state. We believe these new experimental
observations beg for more theoretical and experimental investigation of
frustration driven lattice distortions.

The authors would like to thank P.~Mendels and M.~Horvati\'{c} for helpful
discussion. A.~Keren would like to thank the Israel - U. S. Binational
Science Foundation for supporting this research.


\begin{references}
\bibitem{toulouse}  G. Toulouse, Commun. Phys. {\bf 2}, 115 (1977).

\bibitem{reviews}  For recent reviews see: A.P. Ramirez, Annu. Rev. Mater.
Sci., {\bf 24}, 453, (1994); {\it Magnetic Systems with Competing
Interactions}, edited by H.T. Diep (World Scientific, Singapore, 1994); P.
Schiffer and A.P. Ramirez, Comm. Cond. Mat. Phys., {\bf 18}, 21, (1996).

\bibitem{BinderPMP86}  K.~Binder and A.~P.~Young, Rev. Mod. Phys. {\bf 58},
801 (1986).

\bibitem{dunsiger}  S.R. Dunsiger, R. F. Kiefl, K. H. Chow, B. D Gaulin, M.
J. P. Gingras, J. E. Greedan, A. Keren, K. Kojima, G. M. Luke, W. A.
MacFarlane, N. P. Raju, J. E. Sonier, Y. J. Uemura, and W. D. Wu, Phys. Rev.
B {\bf 54}, 9019 (1996).

\bibitem{gingras_ymoo}  M.J.P. Gingras, C.V. Stager, N.P. Raju, B.D. Gaulin
and J.E. Greedan, Phys. Rev. Lett. {\bf 78}, 947 (1997).

\bibitem{gardner_y2mo2o7}  J.S. Gardner, B.D. Gaulin, S.-H. Lee, C. Broholm,
N.P. Raju, and J.E. Greedan, Phys. Rev. Lett. {\bf 83}, 211 (1999).

\bibitem{gaulin}  B. D. Gaulin,J. N. Reimers, T. E. Mason, J. E. Greedan,
and Z. Tun, Phys. Rev. Lett. {\bf 69}, 3244 (1992).

\bibitem{y2mn2o7}  J.N. Reimers, J. E. Greedan, R. K. Kremer, E. Gmelin, and
M. A. Subramanian, Phys. Rev. B {\bf 43}, 3387 (1991).

%\bibitem{csnicrf6}  M. J. Harris, M. P. Zinkin, Z. Tun, B. M. Wanklyn, and
%I. P. Swainson, Phys. Rev. Lett. {\bf 73}, 189 (1994).

\bibitem{MendelsPRL00}  P. Mendels, A. Keren, L. Limot, M. Mekata, G. Colin,
and M. Horvati\'{c}, Phys. Rev. Lett. {\bf 85}, 3496 (2000).

\bibitem{SCGO}  X. Obradors {\it et al.}, Solid State Commun. {\bf 65}, 189
(1988); A. P. Ramirez, G.P. Espinosa and A.S. Cooper, Phys. Rev. Lett. {\bf %
64}, 2070 (1990).

\bibitem{BSGZCO}  I. S. Hagemann, Q. Huang, X. P. A. Gao, A. P. Ramirez, and
R. J. Cava, Phys. Rev. Lett. {\bf 86}, 894 (2001).

\bibitem{mydosh}  J. A. Mydosh, {\bf Spin Glasses} (Taylor and Francis,
London) 1993.

\bibitem{BoothPRB00}  C. H. Booth, J. S. Gardner, G. H. Kwei, R. H. Heffner,
F. Bridges and M. A. Subramanian, Phys. Rev. B. {\bf 62}, R755 (2000).

\bibitem{RajuPRB92}  N. P. Raju, E. Gmelin and R. K. Kremer, Phys. Rev. B 
{\bf 46}, 5405 (1992).

\bibitem{ClarkRSI95}  W. G. Clark, M. E. Hanson, and F. Lefloch, Rev. Sci.
Instrum {\bf 66}, 2453 (1995).

\bibitem{VillainZPhysB79}  J. Villain, Z. Physik {\bf B} 33, 31 (1979).

\bibitem{TeraoJPSJ95}  K. Terao, JPSJ {\bf 65}, 1413 (1996).

\bibitem{Yamashita}  Y. Yamashita and K. Ueda, Phys. Rev. Let. {\bf 85},
4960 (2000).

\bibitem{Bellier00}  L. Bellier-Castella, M.~J.~P.~Gingras, P.C.W.
Holdsworth and R. Moessner, Cond-Mat 0006306.

\bibitem{MelziPRL00}  R.~Melzi, S.~Aldrovandi, F.~Tedoldi, P.~Carretta,
P.~Millet, F.~Mila, Phys. Rev. Let. {\bf 85}, 1318 (2000).
\end{references}
\end{document}